\documentclass{article}
\usepackage[margin=3cm,a4paper]{geometry} 
\usepackage{amsmath,amsthm} 
\usepackage{amsfonts} 
\usepackage{graphicx}

\newcommand{\eq}{\begin{equation}}
\newcommand{\eqend}{\end{equation}}

\begin{document}

\title{\huge Immunological Approaches to Load Balancing in MIMD Systems}
\author{James J. Clark\\
Department of Electrical and Computer Engineering\\
and Centre for Intelligent Machines\\
McGill University}
\maketitle

\begin{abstract}
Effective utilization of Multiple-Instruction-Multiple-Data (MIMD)
parallel computers requires the application of good load balancing
techniques. In this paper we show that heuristics derived from
observation of complex natural systems, such as the mammalian immune
system, can lead to effective load balancing strategies.
In particular, the immune system processes of regulation, suppression,
tolerance, and memory are seen to be powerful load balancing
mechanisms. 

We provide a detailed example of our approach applied to
parallelization of an image processing task, that of extracting the
circuit design from the images of the layers of a CMOS integrated
circuit. The results of this experiment show that good speedup
characteristics can be obtained when using immune system derived
load balancing strategies.
\end{abstract}

\normalsize
\section{Introduction}
In the quest for increased rates of computation computer scientists
and engineers have turned toward parallel computing systems. Such
systems speed computational tasks by dividing the task into parts and
executing each part in parallel with separate computational engines.
Early parallel computers were of the Single-Instruction-Multiple-Data
(SIMD) form \cite{flynn}, wherein each processing element executes the
same computations on different data. The effectiveness of these types
of parallel computers are limited by the extent to which an arbitrary
task can be divided into a set of identical sub-tasks. There are many
problems which are easily partitioned in this manner, for example
image filtering and other tasks which involve the application of local
operations repeatedly over a field of data. The large majority of
computational tasks, however, are very difficult to partition into
sets of identical sub-tasks. Implementing these tasks in SIMD
computers will result in a very inefficient use of computational
resources.

A more general and powerful class of parallel computers are the
Multiple-Instruction-Multiple-Data computers \cite{flynn}. These
computers consist of a set of processing elements that can execute
different computations on different data. The flexibilty added by the
capability of the processors to execute independent programs permits a
much wider range of tasks to be efficiently parallelized than is the
case for SIMD systems. A completely general MIMD system, one which
allows its constituent processing elements to perform any computation
whatsoever and to access whatever data it requires, presents an
intractable programming or specification problem. The space of
possible programs is vast compared with a SIMD system, and lacks
suitable structure to permit finding the optimal set of programs for a
given task. Typically, the intractability of MIMD programming is
attacked by applying constraints, usually architectural, to the space
of possible programs. These constraints reduce the space of the
possible programs, and also provide a template to guide the programmer
in the specification of a program for a given task. 

Besides the difficulty in constraining the programming, developers of
algorithms to be implemented on MIMD machines are faced with another
difficult problem, that of {\em load balancing}.
In any multi-processor system the most important measure of
performance is the speedup in completing a given task obtained by
using a number of processors over using a single processor. If a
processor is performing non-useful work (i.e. work not relevant to
solving the problem at hand), or is duplicating relevant work being
done by another processor, then it is not contributing to speeding up
the task. It is clear that what needs to be done in order to achieve
the maximum possible speedup is to distribute the workload amongst the
processors as evenly as possible. This process is called load
balancing, and is one of the most important facets of MIMD system
design \cite{ahmad}. 

One can make the observation that many natural systems, such as the
mammalian immune system, can be thought of as MIMD systems. These
natural systems have been ``programmed'' by evolution to carry out many
types of tasks and provide examples of efficient algorithms that can
be adapted to other MIMD systems that exhibit similar characteristics.
Others have made this observation before (e.g. Paton \cite{paton}),
but have emphasised the use of computational metaphors and models in
forming biological models. We want to turn this around, and
concentrate on the application of biological metaphors and models in
specifying computational processes. Specifically, we are interested in
abstracting the ``load balancing'' strategies that are in effect being
employed by these natural systems. Primary amongst these strategies
are the {\em regulatory} processes that play a crucial role in immune
system function. We will also examine the role that other immune
processes such as immunological memory have to play in load balancing.

We end the paper with an application of our approach to the
parallelization of an image processing task, that of extracting an
electrical circuit netlist from images of the IC fabrication process
masks. This problem is difficult to parallelize due to the many serial
sub-problems that must be solved. In spite of this, our approach
yields respectable speedup performance. The population dynamics that
arise in these simulations are clearly similar to those in immune
systems and illustrate the action of the various load balancing
strategies that are in use.

\subsection{Load Balancing in the Immune System}

Load balancing is one of the most important and difficult aspects of
programming of MIMD computers. In brief, it is the allocation of
resources amongst the active elements of the system in a way which
leads to the fastest execution of the task at hand.  We propose to
approach the problem of load balancing by using insights gained from
observing natural systems, such as the immune system.

The immune system \cite{jerne,kuby} is a massively parallel MIMD system. It
consists of about $10^10$ cells, each of which have different roles to
play in its functioning. The immune system has evolved to respond
effectively and quickly to appropriate stimuli, and to make efficient
use of its constituents. Thus it is a good example from which to
abstract principles of effective load balancing.

Some of the primary attributes of the immune system are:
\begin{itemize}
\item Recognition of environmental conditions
\item Tolerance to irrelevant conditions
\item Response speedup through immunological memory
\item Response regulation 
\item Dynamic cell population control
\item Message passing via a shared environment
\end{itemize}

One important lesson we can learn from the immune system with respect
to load balancing is that cellular populations are dynamic, and
populations of a given cell type are increased or decreased only when
appropriate conditions are detected. These conditions are either
signaled directly, through sensing of environmental factors (such as
presence of antigens) or indirectly through messages sent by other
cells.  Likewise, load balancing schemes are classified as being
either static or dynamic. In static load balancing, the schedule of
tasks that a given agent will perform is specified before execution
begins. This requires detailed {\em a priori} knowledge about the
characteristics of the tasks needed in the course of executing the
application. If the precise nature of the application is not known
beforehand, static schedules will generally be non-optimal. Dynamic
load balancing is a form of load balancing that is more efficient in
the face of incomplete a priori information about task
characteristics. In dynamic load balancing, the assignment of tasks to
processors (or agents) is done based on the current state of the
system. Clearly, any load balancing scheme based on the immune system
metaphor will be dynamic.

Let us now examine each of the attributes of the immune system listed
above and discuss their application to load balancing.

The recognition capability of the immune system is a common
requirement of computational tasks. It is often the case that a
program must carry out one or more searches for a given pattern. In
addition, the task usually requires that some action be carried out
once the search has found a target. In itself, recognition is not a
process which serves load balancing ends. Rather, it is a process
which must itself be a target of load balancing
processes. Nonetheless, recognition processes, by their essentially
parallel nature, put constraints on the load balancing processes. In
general, speeds for search operations scale linearly with the number
of units engaged in the search, and the immune system is no different
in this regard. This does not mean, however, that the optimal strategy
is to employ all resources in the search activity and to redirect all
these resources to other activities once the target has been
found. The system must be able to respond quickly to unknown
conditions, and assignment of all system resources to a given task may
slow response to some events.

Tolerance of ``self-antigens'' is crucial in the immune system, as
otherwise the immune system would attack everything in the body, and
not limit itself to external invaders. In load balancing, tolerance
can be viewed as limiting irrelevant processing. That is, there may be
environmental conditions which signal or suggest that a certain
computation should be done, but a computation which does not
contribute to the task being carried out. In this case the
environmental condition, or its signal, should be ignored, or
tolerated, by the system. In the context of load balancing in MIMD
systems, the symptoms of the analog to an autoimmune disease would be
the slowing down of the rate in which problems are solved.

In the immune system tolerance is produced in a number of ways.  The
most well known is that of clonal deletion, wherein a T-cell
responsive to a self-antigen is killed in the thymus if it has
responded. This happens early in life, so there is a chance that
self-antigens which appear later in life (e.g. during puberty) will
cause immune responses. For these case, the so-called peripheral T
cell tolerance mechanisms come into play. In peripheral tolerance,
auto-immune responsive cells are not killed but instead inactivated, a
process referred to as {\em anergy} \cite{kuby}. It is thought that
anergy results from a lack of proper co-stimulation of the cell. In
co-stimulation, the presence of the self-antigen must coincide with
the presentation of B7 membrane proteins by Antigen Presenting Cells
(APCs) if there is to be an effective response. Anergy may also be
caused if a variant form of the antigen is encountered. Instead of
anergy, tolerance can be mediated by suppression from other
lymphocytes (called suppressor T-cells). The mechanisms by which
suppressor T-cells are activated are similar to those thought to
induce anergy.  Anergy is not irreversible. The cytokine IL-2 can
cause anergic T cells to become reactivated.

The immune system employs a number of strategies for optimizing
response speeds. One of these is {\em immunological memory}
\cite{sprent}. When a given condition is first encountered, an
effective response to it may be slow in occuring. If a small number of
units are henceforth dedicated to detecting this specific condition,
then future responses may be much faster. This is actually a load
balancing strategy as it dictates that a small number of agents be
assigned to carry out a seemingly irrelevant task in the hope that a
rapid response can be attained when environmental conditions change
appropriately. In the immune system T-cells can either be ``effector
cells'', capable of taking direct action against invaders, or ``memory
cells''. A recent theory of T-cell genesis \cite{farber} holds that
low antigen levels can cause young, naive, T-cells to become memory
cells instead of effector cells. In this case, then, environmental
conditions can determine the function of a given agent. Another theory
\cite{sprent} holds that effector T-cells become memory cells after a
certain number of cell divisions. In load balancing terms, this
approach is saying that a T-cell that has undergone many divisions in
the past has done so in reponse to relevant environmental conditions,
which are likely to arise again in the future, even if they are
currently not present.

An important way in which responses can be sped up is to use positive
feedback. In the immune system T4 ``helper'' cells stimulate
themselves to reproduce through release of chemical messages (proteins
known as cytokines). Because of this positive feedback, the population
of T4 cells can increase very rapidly. Unchecked, however, the
positive feedback would cause the T4 cell population to grow until
resources become used up. This feedback is limited somewhat by the
fact that the cytokines typically have a short lifespan.
Another way to prevent such a runaway is to
provide a counteracting negative feedback. One does not want this
negative feedback to simply cancel out the positive feedback,
otherwise no speedup would be attained. A way out of this dilemma is
to delay the negative feedback. This is done in the immune system with
a two stage regulatory process, in which the T4 cells facilitate
increases in the population of T8 (cytotoxic) cells which then
suppress the population of T4 helper cells. The suppression does not
become significant until the T8 population grows, however, leading to
a delay in the response suppression. In this way, a rapid response is
assured, but one which does not run away.

Two-stage regulation can also be used to choose between possible
responses, by suppressing the inappropriate (or irrelevant)
responses. In the immune system, two classes of helper cells, Th1 and
Th2 operate in this fashion \cite{glimcher}. The immune system has two
basic approaches to deal with an invading antigen. The first is a
humoral, or antibody-mediated immunity (AMI) which is effective
against extracellular antigens. It is triggered by the presence of
antibodies againts a particular antigen.  The other is cell-mediated
immunity (CMI) which is effective against intracellular antigens. It
is triggered by cytokines (chemical messengers) released by various
immune system cells, such as macrophages. A stressed macrophage
releases the cytokine IL12 which, in combination with IL2 enhances Th1
production. Th1 cells in turn secrete IFN- which suppresses Th2
production. This shifts the balance of the immune system towards a CMI
response. On the other hand, antigen stimulated B-cells secrete the
cytokine IL10 which, in conjunction with IL4, suppresses Th1 cells
(through inhibition of secretion of IL12 by macrophages). Th2 cells
also secrete IL4 which further inhibits Th1 production. In this way
the immune system is directed towards a humoral response. As noted by
Abbas {\em et al}, T-cells select one of multiple modes of reactivity,
and the immunological ``self'' is not defined by the repertoire of
T-cell types, but rather by the behaviour of the system in response to
environmental conditions.

\subsection{Redundancy and Irrelevancy Minimization}

Our goal is to abstract from the functioning of the immune system
various principles which can be used in developing load balancing
strategies for MIMD systems.

To this end, we propose that load balancing in the immune system (and
in complex natural systems in general) is best thought of in terms of
as a process for minimizing agent {\em redundancy} and {\em
irrelevancy}. Redundancy refers to the condition wherein an agent is
duplicating work being performed by another agent.  Redundancy is a
problem when there a large number of agents in a region compared with
the number needed to carry out a task. One must be careful to
correctly determine this number. For example, in searching for an
object randomly placed in a region, the more agents that are employed,
the sooner the search will be completed. Once the object has been
found, however, an effector operation may only require a few
agents. Hence redundancy is a problem for the effector operation,
whereas it was not a problem for the search task. 
In the immune system most tasks, such as detection and elimination of
antigen, are essentially parallel in nature, so redundancy resolution
is not so important as in some other complex systems. 

Irrelevancy, on the other hand, refers to the condition wherein an
agent is performing work which is not contributing anything to the
overall task of the system. Irrelevancy is a problem when there are a
small to moderate number of agents relative to the number required to
carry out a task.

We propose that general load balancing strategies should aim to
minimize both redundancy and irrelevancy. Such load balancing
strategies will neccessarily involve both the {\em detection} and {\em
correction} of redundancy and irrelevance. Redundancy will usually be
easy to detect, merely by the presence of other agents performing the
same tasks. The major problem is to determine whether or not these
other agents are actually neccessary, as in the case of a parallel
task such as search.

Some techniques for detecting and alleviating redundancy that are
found in nature (and in the immune system) are:
\begin{itemize}
\item {\em Dominance} The dominant agent gets to perform a certain activity while
other agents are inhibited from doing so. The dominance can be based
on fixed agent characteristics, such as an identification code, or can
be based on variable agent attributes, such as age, time spent
performing a task, or value of an internal state variable
(e.g. health). The dominance could even be based on the outcome of a
random event (coin flip). 
\item {\em Suppression}. Agents change their behaviour when there are
many neighbors of the same type. In the immune system this does not
occur, as most types of cells are auto-stimulatory. 
\item {\em Change Inhibition}. An agent can write environmental messages that
inhibit other agents from changing to its type or behaviour. Again, in
the immune system this type of process is not common.
\item {\em Diffusion}. Due to diffusion, agents of the same type will
tend to move away from each other. Anti-tropism to secrete chemical
messages could also be used to repel agents from other like agents.
\item {\em Mutation}. The specificity of agents, and hence their
behaviour can change due to random mutations. Thus a group of agents
that initially have the same behaviour will change over time to have a
range of different behaviours. This effect is found in the activity of
antibodies in the immune system. In order to wide range of different
antigens using limited antibody resources it is crucial that
redundancy in reactivity to antigen be minimized.
\end{itemize}

Irrelevancy is difficult to detect, let alone correct. Some
possible strategies for detection of irrelevancy are: 
\begin{itemize}
\item Lack of similar agent types in the neighborhood. In this immune
system, the presence of similar agent types is detected through
cytokines released by these agents. Lack of these chemical messages
can suppress a cell's activity or lead to anergy.
\item Lack of expected environmental conditions or appropriate
messages from other agents. Again, in the immune system, lack of
expected conditions can lead to suppression or anergy.
\item Presence of unexpected environmental conditions or inappropriate
messages from other agents. In the immune system, presence of certain
cytokines can also lead to suppression or anergy.
\item The value of a dynamic internal state variable exceeding a
threshold. For example, a timer could be used to 
measure the time since an agent began a task, such as searching for
some environmental condition. In the immune system, for example, it is
thought \cite{sprent} that once T-cells have divided a certain number
of times they change their behaviour from effector cells to that of
memory cells.
\item Lack of similar agent types in the neighborhood. In this immune
system, the presence of similar agent types is detected through
cytokines released by these agents. Lack of these chemical messages
can suppress a cell's activity or lead to anergy.
\end{itemize}

Strategies for correction of irrelevancy include:
\begin{itemize}
\item Change an agent's type to the majority type of its
neighbors. The reasoning behind this approach is that the majority of
the neighbors are likely to be performing relevant work. 
\item Change back to the agent's previous type, if there was one. This
is a reset, in the hope that the previous type will be better suited
to correcting its mistake. This assumes that the agent knows the type
of its predecessor, and that the agent has a predecessor. 
\item The agent can change its motion law to escape the region that it
currently is in. 
\item Change to another agent type. This should not be
done randomly, but based on some rules perhaps modulated by
environmental conditions. 
\end{itemize}

In our view of load balancing the goal is to convert agents engaged in
irrelevant or redundant activities to types that do useful
work. However, in order to heighten responsiveness to new conditions
in a dynamic environment, it may be advantageous to have agents
performing apparently non-productive tasks. For example, in the
application of finding, linking, and tracking edges in a time-varying
image, agents that are searching for edges in regions of the image not
currently populated by edges could be prevented from changing their
activity to a more productive task. Then, when the image changes with
time to a state where there were now edges in the area the
non-productive agents are searching in, these agents will quickly find
the edges. The time taken for these agents to find the new edges is
likely to be less than the time required if these agents were doing
some other task, or had moved to another region of the data
space. Thus, we can employ the strategy of immunological memory. The
number of agents carrying a particular type of ``immunological
memory'' need not be large, if the immunological enhancement described
above is present. Hence the loss of performance due to the
non-productive agents is small, and is offset by the increased speed
of response to novel stimuli.

\subsection{Regulatory Processes}

In a complex system, in a complex environment, tasks are rarely simple
and unitary. Often there are many subtasks to be worked on in
parallel. This means that what may seem to be an irrelevant activity
for one subtask is relevant for another subtask. Thus, load balancing
algorithms based on irrelevancy minimization must be careful to not go
too far. In particular, irrelevancy minimization strategies must be
prevented from allocating all processing to a single activity. In the
immune system this control is attained through the use of {\em
regulation}.  Regulation is an important concept that has not yet
found its way into the parallel processing field. Regulation is a way
to control resource utilization. If there are a number of tasks to be
performed by a MIMD system (e.g. searching for different types of
features in an image) and only there are only a relatively small
number of processors, then there arises the question of how the
processors should be allotted to the various tasks. Regulation handles
this problem by having processors facilitate the action of some types
of processors, while inhibiting or suppressing the activity of
others. Positive feedback, arising from self-excitatory or
auto-catalytic behaviour permits a rapid response to suitable
conditions. Left unchecked, however, this feedback will cause the
system to allocate more and more resources to this response,
eventually using up all the resources.

In the immune system many cell types exhibit auto-stimulation
behaviour. This allows a rapid response to be mounted to various
stimuli. This response is regulated, however, through use of delayed
negative feedback, mediated by other cell types. The delay permits the
response to initially rise very quickly, but later to be reduced.
In the earlier discussion of the immune system function we saw that
such a positive-negative feedback configuration exists between Th1 and
Th2 helper T-cells.

It is our contention that the implementation of parallel computer
programs for normally poorly parallizable tasks can potentially be
improved with this notion of regulation, or control of resource
utilization, through processor interaction.

Regulatory processes such as the one described above can lead to
oscillations if not carefully controlled. Likewise, limit cycles in
cell or agent populations of various types can result from some load
balancing strategies. This is especially likely to happen as a result
of the redundancy/irrelevancy dichotomy. For example one can obtain a
limit cycle, or oscillation, if one alleviates redundancy by changing
agent behaviour if the agents neighbors are of the same type
(self-suppression), followed by alleviating irrelevancy by changing
agent behaviour to that of its neighbors. Such limit cycles are
commonly observed in biological systems. For example, limit cycles in
tumor cell and lymphocyte populations have been observed
\cite{kirschner}.  There are a number of strategies for preventing
limit cycles.  Cell or agent diffusion can disrupt them. Another
approach is to never allow an agent to change its behaviour to one of
its parent or ancestor types. One can also suppress (but not disallow)
such changes, thereby dampening out incipient limit cycles.

\section{The MIMD Model and Programming Methodology}
As alluded to in the introduction, one of the difficult aspects of
programming for MIMD systems is the need to constrain the form of the
program. In this section we present a general computational template
that specifies the form of the MIMD systems that we will restrict the
programmer to follow. We will use this particular model to specify
MIMD programs that demonstrate the utilization of the immunological
load balancing principles described in the previous section.  We
believe that our programming methodology will be applicable to a wide
range of shared-memory class MIMD machines, and its usage will yield
benefits in terms of parallelizability of algorithms and in terms of
the programming effort required.

\subsection{A Model Framework for MIMD Programs}
The model that we use is the one developed by Hewes \cite{hewes}.
In this model, the constraints on the allowable MIMD systems are both
explicit and implicit. The explicit constraints on the MIMD systems
manifest in the form of a program template. This template dictates, in
a quite restrictive manner, the nature of the allowable programs,
while retaining enough freedom to permit a wide range of programs to
be implemented.

In what follows, we refer to each processor in the MIMD system as an
{\em agent}.  The first way in which we constrain the program is to
require that the program being executed by each agent be represented
by what we call the {\em Agent Characteristic}. The agent
characteristic is an object that consists of descriptions of five
subprogram objects: $Pr; Pw; Pa; Pu; Pm$.  Different agents that have
the same characteristic are said to have the same agent {\em type}.
The subprogram descriptions are converted to executable programs and
then executed, in a sequential and cyclic manner, by the agent's
computational hardware.  The sequence of operations is $Pr$ (read),
$Pu$ (state update), $Pw$ (write), $Pa$ (agent alteration), and $Pm$
(move). The operations in the processing cycle are performed
serially. Once the processor has reached the end of a processing
cycle, i.e. once it has finished the move operation, it repeats the
cycle, beginning with the read operation.

Another important constraint in this model is that the programmer
must specify a finite set of {\em predefined} agent
characteristics. Each processor can only use characteristics that are
in this set. In effect, the MIMD ``program'' produced by our approach
is just this set plus the initial association of characteristics
from this set to each individual agent.

The five subprograms that make up the Agent Characteristic 
have a restricted functionality, and have the following interpretations:
\begin{itemize}
\item {\em Agent State Update}, $Pu$: The subprogram $Pu$ is a state
update operation, mapping the current state into a new state.  Whereas
the characteristic of an agent encodes its "static'' aspects, its state
encodes its "variable'' aspects.  Different agents of the same type can
have different state. In the programming template the state is left
unconstrained. It can be as simple or elaborate as the programmer
desires.
\item {\em Agent Read Operation}, $Pr$: The subprogram $Pr$ is
interpreted in our model as a read or sense operation. It changes the
value of data objects in the agent state according to the values of
data objects in the agent's {\em receptive field}.  The receptive
field of an agent is a description of the parts of the shared memory
space that the agent has read and write access to. 
\item {\em Agent Write Operation}, $Pw$: The subprogram $Pw$ is
interpreted as a write or effector operation. It changes the value of
data objects in the agent's receptive field according to the value of
data objects in the agent state.
\item {\em Agent Movement Operation}, $Pm$: The subprogram $Pm$ is
interpreted as an agent motion operation. Given the current state and
receptive field of the agent this subprogram produces a new receptive
field.
\item {\em Agent Alteration Operation}, $Pa$: The subprogram $Pa$, is
an agent type alteration operation, which changes the characteristic
of the agent according to the current state of the agent. Recall that
the set of possible agent characteristics is predetermined.
Application of this subprogram can result in a change of the agent
type. The alteration operation can also alter the agent state. If
there is no change in the agent characteristic, this change in state
would be redundant with that caused by the agent's state update
operation. If however, the agent changed it's characteristic, this
change in state can be used to initialize the ``new'' agent's state.
\end{itemize}

Whereas individual complex units are very diverse and localized, the
environment will be assumed to be homogeneous in its properties and
distributed. The environment acts as a stage on which the complex
units act out their behaviors, and mediate their interactions. 
Thus, we will assume that the MIMD system being programmed is of the
shared memory variety, or a so-called ``blackboard'' system.

The approach outlined above is intended to provide constraint to the
programmer of MIMD systems, but also to follow the general features of
natural complex systems, such as the immune system.

The key aspect of this framework which permits load balancing to be
carried out is the Agent Alteration subprogram. This subprogram allows
an agent to change its characteristic in response to internal or
external (environmental) conditions. Agents can also read and write
the shared memory. In this way different agents can interact with each
other. The interaction capability of the agents in our approach leads
directly to cooperativity, as well as to facilitation (as in
autocatalytic networks) or suppression (regulation) of program
functioning. 

\subsection{Programming Methodology}
The programming of MIMD systems is a difficult and sometimes
intractable problem. We have attacked this difficulty by providing
constraints on the programming process. In addition to the programming
template described in the previous section, we must also provide a
programming {\em methodology} to guide programmers in how to best use
the template.

The programming methodology involves a number of steps that the
programmer should carry out in developing a program to solve a give
problem.  The first step is to break the overall task into a number of
distinct phases or sub-tasks. Different agent types are developed to
carry out the primary activities in each of these phases. The Agent
Motion, State Update, Read, and Write subprograms are programmed so as
to result in the desired behaviour needed to carry out these
activities.

Up to this point, the programmer has only needed to be concerned with
how to carry out a given task. Now the programmer must incorporate
load balancing strategies. In our framework, this is done by embedding
the dynamic load balancing strategies described earlier in the paper
and should include regulatory processes to speed up response times.
These load balancing behaviours are mainly controlled through
specification of the Agent Alteration subprograms.

If needed, additional agent types can be defined to act as
catalysts or helpers for the other agent types. The need for these
agent types can be deduced through observation of hot-spots or
bottlenecks in simulations. This is analogous to the function of
helper-T cells in the immune system.

In contrast to natural systems, such as the immune system, where agent
behaviours are developed through evolution, MIMD programmers must rely
on analyses and examination of simulation results to optimize agent processing.
Simulation is useful in tuning the various agent characteristics
to speed up sub-task execution or in avoiding freeze-ups, deadlocks or
limit cycles. Speedup analyses are also useful in gauging the
effectiveness of various load balancing strategies. These analyses are
done by executing multiple runs with varying numbers of agents, and
with randomly selected initial agent spatial distributions (and/or
agent state). Termination times are noted and tabulated against number
of agents.

\section{An Example: VLSI Layout Extraction}
In this section we describe an example of how to use our programming
template and methodology to implement, in a MIMD computer, a problem
that is difficult to parallelize. We will apply the load balancing
strategies that we have derived from examination of immune system
functioning. Our approach to programming of MIMD systems is quite
different than that used with standard programming languages.

The problem we consider is that of extracting a circuit netlist from a
fabrication mask level description of a VLSI layout. As we will see,
carrying out the layout extraction task requires execution of both
parallel and serial subtasks, making efficient implementation on a
parallel computer difficult.

The VLSI layout extraction process \cite{vlsi} produces a circuit
netlist, which is a listing of circuit elements and their
connectivity, from a description of the masks used in the fabrication
of the integrated circuit. For the purposes of describing the
application of our MIMD programming approach to VLSI layout extraction
we take a simplified view of the {\em layout} of a CMOS VLSI circuit. A
{\em layout} is a graphical description of the physical instantiation of the
circuit, composed of rectangles uniformly colored with various values
corresponding to the different physical fabrication masks. In the
simplified view the different masks, or layers as we will call them,
used in the fabrication process
are Metal1, Metal2, Poly (polysilicon), Diff (diffusion), PSEL (doping
select), and Contact. The two metal layers are used to form wires that
connect between transistor elements. They are insulated from each
other. The poly layer is used both as a wire
(insulated from the metal layers) and to form the gate regions of
transistors. The diff layer is used both as a
wire (insulated from the metal layers) and as the source and drain
regions of transistor elements. The diff and poly layers are not
insulated from each other and form a transistor when they overlap. In
the graphical representation we use, overlapping a diff wire with a
poly wire causes the diff wire to split into two separate pieces, on
either side of the poly wire. These form the source wire and the drain
wire. The poly wire is left alone. The PSEL layer determines the
polarity of the transistor. Contact regions cause a connection to be
formed between the first metal layer (metal1) and any other layer that
overlaps the metal1 layer in the contact area. The design rules permit
only two layers at a time to be connected with a contact, and no
direct contact can be formed between a poly region and a diff
region. In our simple model, we will assume that the only elements
contained in the circuit are transistors and wires connecting them.

The job of the layout extractor program is to detect the presence of every
transistor in the layout and determine the connectivity between
them. As input to our extractor program we take a rasterized
representation of the layer masks. This is a bitmap with the number of
bit planes equal to the number of layers in the IC fabrication process
(six in the current simplified model). If the bit in a particular bit
plane at a location in the bitmap is high then the corresponding layer
is present at that location, otherwise it is absent. The output of the
program is a series of statements, each having one of the two
following syntactic forms:
\begin{itemize}
\item FET id number: S - node number, D - node number, G - node
number, L - length, W - width, Time = time steps
\item Contact id number: Node node number == Node node number, Time = time steps
\end{itemize}
where FET ::= NFET  or PFET  and the other values are all
integers. The first type of statement expresses the presence of a
transistor (either an NFET (n-channel MOSFET) or a PFET (pchannel
MOSFET)), along with the node numbers for its source, drain, and gate
terminals. The ``Time'' value is the number of time steps the program
took to determine the node numbers for the device. The second type of
statement describes equivalences between different node numbers. Our
algorithm assigns a unique node number to each contiguous region on a
single layer. Different nodes may be connected electrically through a
contact, however. Thus, our algorithm outputs an equivalence statement
for each contact in the circuit, describing the two different nodes
that are electrically connected by the contact. Note that, in many
layouts, multiple contacts are used to connect a pair of nodes. Our
algorithm will output redundant equivalence statements in this case. 

As an example of the type of input that the system sees, we show in
figure 1 the layout of a simple four-transistor CMOS digital NAND
circuit. The circuit contains eighteen contacts. Overlaying the layout
are symbols indicating the location of a set of agents of various
types at a time midway through the execution of the extraction
process.
\begin{figure}
\begin{center}
\includegraphics[width=3in]{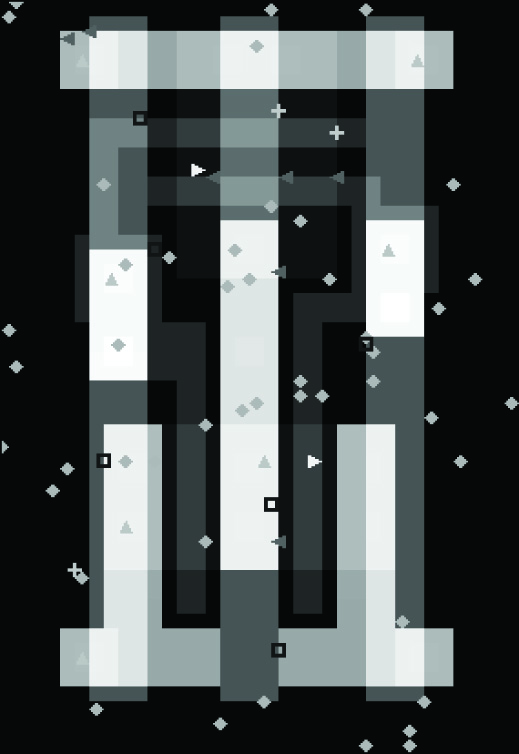}
\end{center}
\caption{The layout for a CMOS NAND circuit, displayed with the
agents during a simulation of the parallel extraction
algorithm. (diamond - layer finder, left-triangle - node labeller,
right-triangle - fet labeller, down-triangle - fet output, up-triangle
- contact finder, square - node director, plus node propagator)}
\end{figure}

\subsection{MIMD Layout Extraction Algorithm}
There are many MIMD algorithms that could conceivably be written to
solve the VLSI layout extraction problem. We will present one such
algorithm here, but make no claims as to its optimality. 

Following the general programming methodology outlined earlier, we
break the programming task into two steps. First we determine the
sequence of sub-tasks that need to be performed and develop specific
agent {\em characteristics} to carry out each sub-task. Then we add
load-balancing functionality to each agent type, as well as create
additional agent types that solely carry out load balancing
activities. The load-balancing behaviours that we implement are based
on the immunological metaphor as discussed earlier in the paper.

We can break the layout extraction problem down into a number of
subtasks as follows: 
\begin{itemize}
\item 1) Search for wire boundaries. 
\item 2) Trace and label wire boundaries such that each wire has
only one label and each label is attached to at most one wire. 
\item 3) Search for transistors along diff wires. 
\item 4) For each transistor, trace the transistor
boundary to measure the gate length and width and to compile the
labels for the gate, source, and drain terminals. Once the trace is
complete, output a transistor description statement. 
\item 5) Search for contact areas. For each contact, determine the labels of
the wires that overlap the contact area, and output a node equivalence statement.
\end{itemize}
Note that some of these tasks are inherently parallel (the search tasks), while the
others are inherently serial (the labelling and tracing tasks). 
The strategies that a given agent would use for load balancing would
depend on whether the agent was searching or labelling.

To carry out these five subtasks, our algorithm employs the following
types of agents, layer finder, node labeller, fet labeller, fet
output, contact finder. Two additional agent types, node director, and
node propagator, do not directly contribute to the task, but are
solely concerned with load-balancing. These are helper types, similar
in abstract function to helper cells in the immune system.

The behaviour of the five task-oriented agent types is summarized in the following
paragraphs.

{\em Layer Finder}: The task of agents of the layer finder type is to
search the environment for the boundaries of unlabelled wires of
metal, poly, or diff layers. The search pattern used is a raster
scan. 

{\em Node Labeller}: The job of the node labeller is to move along the
boundary of the wire, while writing a unique label to the shared memory along
the wire's boundary.

{\em Fet Labeller}: The fet labellers trace the DIFF layer wire
boundaries and search for the combination of DIFF and POLY layers that
indicate the presence of a transistor. If such a coincidence is found,
a unique fet label is written to the environment at that point.

{\em Fet Output}: The role of the fet output agents is to measure the
length and width of the gate region of the transistor, and to write the
labels on the associated POLY and DIFF layers to a file, along with
the transistor gate length and width.

{\em Contact Finder}: The contact finder's role is to sit and wait at
a contact location until the wires overlapping the contact area become
labelled. Once the wires have been labelled, the contact finder agent
outputs the labels of the two wires to a file in the form of an
equivalence statement.

\subsection{Load Balancing Strategies}

Some aspects of the behaviour of each agent type are directed towards
load balancing. These are listed below, along with their
interpretation with respect to the immunlogical metaphor. The final
two agent types in this list are concerned only with load balancing.
They do not directly contribute to execution of the layout extraction
task.

{\em Layer Finder}: If the layer finder agent detects an already labelled
boundary, that agent alters it's type to that of a node labeller
agent. The detection of the labelled edge is taken to be an indication
that the agent is {\em redundant}. In the immune system the presence
of a certain cytokine (analogous to the boundary label here) can
suppress the proliferation of one type of cell, and enhance that of
another (analogous to the suppression of layer finder types and
enhancement of node labeller types).

The load balancing with respect to the layer finder agents must be
approached with care. If layer finder agents are changed
to other types too soon, appropriate conditions for their proper
function will not be present and  these changed agents can be
considered as irrelevant. Thus, this load balancing should be
delayed somehow until the conditions are such that large numbers of
layer finders are no longer required. One approach by which this can
be done automatically is that of multi-stage delayed suppression. In
this technique, the activity of one type of agent (in this case the
layer finders) causes the activity of some other type(s) of agent(s)
(in this case the node labellers) which in turn activates a third set
of agents. This third stage of agent activity then inhibits (by
causing a change in type) the first type of agent. The advantage of
this form of suppression is that the effect is delayed until the
appearance of the third stage of agents, at which time the processing
of the first type of agent is not needed. 

The multi-stage suppression is carried out by the node director and
node propagator agents, which are two generations removed from the
layer finders (that is, these types of agents cannot be created from
layer finders, but only from descendents of the layer finder 
agents). A layer finder agent is caused to turn into a node propagator
if a node propagator agent is in the agent's receptive field at the
same time a label written by a node director agent is also in the
agent's receptive field.

{\em Node Labeller}: The node boundary labelling process is made
complicated by the requirement that the label be unique, that this
particular label be assigned only to this wire and to no other
wire. The uniqueness of the label can be assured if we use the agent
ID number as the label and if we ensure that a node labeller that
completes the labelling of a wire, as well as any of its descendents
(i.e. agents with the same ID number), cannot change into a layer
finder. This is a strategy for reducing redundancy in the most extreme
sense, as we must ensure that only one agent carries out the node
labelling process on a given node. The uniqueness of response is
similar to the extreme specificity of antibody response to antigen in
the immune system.

{\em Fet Labeller}: The fet label is used to indicate that the node
label for the DIFF wire is stable and can be used by the fet output
agents in composing the transistor output statements. In general, this
is a synchronization activity, in which the execution of a given task
is held until an appropriate signal is given. The analogous process
in the immune system is the release of a cytokine which signals
another type of cell that it is now appropriate to carry out a certain
activity.

{\em Fet Output}: The fet output agent waits for the labels written to
the DIFF and POLY wires on its current boundary to become stable, at
which time the fet information is written out to a file. The agent
then changes to a node propagator type of agent. The waiting activity
is a synchronization process similar to that in the Fet Labeller. 

{\em Contact Finder}: If the contact finder is in a contact region
that has already been captured by another contact finder, or when it
has output a node equivalence statement, it turns into a node
propagator agent. These are redundancy reduction activities. In
general, redundancy reduction of this sort operates by waiting for an
appropriate environmental signal (e.g. a suitable cytokine in the
immune system) followed by a change in the population dynamics
(suppression of one species along with an enhancement of another).

{\em Node Propagator}: The node propagator is a helper agent type. The
purpose of the node propagator agent is to propagate the label written
on the boundary of a wire into the interior of the wire. This filling
in of the wire interior is neccessary for finding contact areas since
contacts are most often found in the interior of wires.  Since the
extraction program must output a node equivalence statement for each
contact, consisting of the node labels for the two wires that overlap
the contact area, the node labels must be propagated to the contact
area.  The motion of the node propagator agents is random when there
are no node director labels to propagate, or where there are no
regions to propagate these labels to. Otherwise the motion is towards
unlabelled areas connected to the labelled areas.  While propagating
the wire lables, the node propagators also check for unlabelled
contact areas. If such an area is found, the agent turns into a
contact finder agent.  Node propagator agents also look for transistor
regions that have been marked by fet labeller agents but not yet been
examined by fet output agents. If such a transistor is found, the node
propagator turns into a fet output agent.

Node propagator agents are rather more complicated than the other
types of agents, but primarily play a communicative role. In this
sense they are like antigen presenting cells in the immune system,
which, through diffusion or other means, communicate the presence of
antigens to effector T-cells. For example, follicular dendritic cells
capture antigen, and then travel to lymphoid organs, where they
interact with effector T-cells.

{\em Node Director}: The node director is a helper agent type. Its
purpose is to retrace the boundary of a traced wire. This retracement
is necessary to signal the node propagator agents to fill in the wire
interior areas with the wire label. This filling in of the wire
interior cannot begin until the wire boundary has been completely
traced, as it is not until that time that a label on any point on the
wire boundary is guaranteed to be the unique node label for that
wire. Thus its role is primarily one of communication.  To ensure
uniqueness of the label, there should only be one node director on
every wire. This is assured by the fact that they are created from the
dominant node labeller agents on each wire. When the node director
agent completes it's retrace it turns into a node propagator agent.

\subsection{Agent Interactions}

In analysing the dynamic behaviour of complex systems such as the
immune system, it is often useful to look at how different types of
system entities interact. Generally, one is concerned with population
dynamics, that is, how the concentration of various entities varies
over time. Thus, the types of interactions that are of interest are
those which facilitate or inhibit the growth of a given species of
system entity. In our MIMD model, such interactions can be direct,
causing one type of agent to turn into another type of agent, or
indirect, merely modulating the probability with which such a
transition will occur.

The agent interactions in the layout extractor algorithm are summarized in
figure 2. The straight lines indicate possible changes of one type
into another. The curved lines indicate the effects agents of various
types have on the probability of a given transitions, either
facilitating or inhibiting the changing from one specific agent type to another. 
\begin{figure}
\begin{center}
\includegraphics[width=5in]{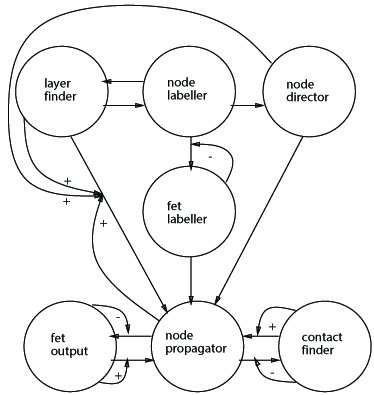}
\end{center}
\caption{Interaction between agents of different types. Straight
lines indicate transitions between agent types. Curved lines indicate
the facilitatory (indicated by a '+') or inhibitory (indicated by a
'-') effect of different agents on the agent transitions.}
\end{figure}
It is instructive to compare this diagram with similar diagrams found
in papers describing the dynamics of the immune system
(e.g. \cite{roitt}). In each case, the populations of various cell or
agent types are enhanced or suppressed by the presence of other cell
or agent types. It is these population interactions in the immune
system that gives rise to its fast and effective response, and which
should lead to effective load balancing in artificial MIMD systems.

In figure 3 we show a population graph of the number of each type of
agent as time progresses for a single run on the layout of a D-flipflop
circuit containing 32 transistors and 120 contacts. In this plot one
can see the dynamics of the load balancing aspect of the algorithm. To
produce this population graph we used the Swarm system from the Santa
Fe Institute \cite{swarm} to simulate the MIMD computer which could
implement our approach. Swarm provides an experimental testbed for
demonstrating and analyzing the performance of solutions to
computational problems that utilize our MIMD programming framework.

\begin{figure}
\begin{center}
\includegraphics[width=5in]{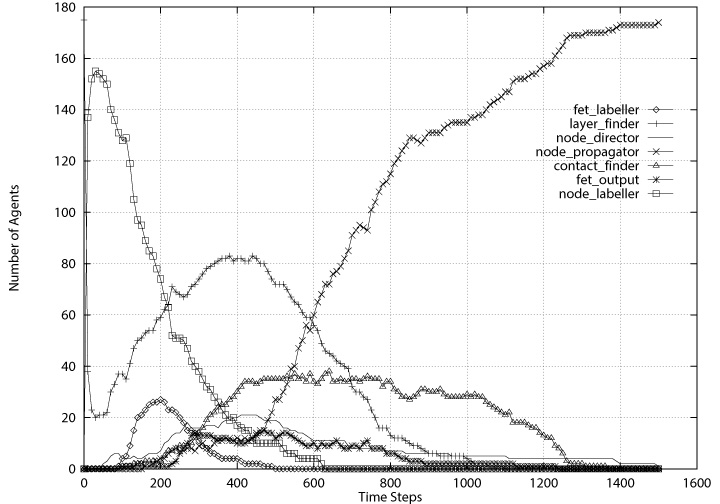}
\end{center}
\caption{The number of each type of agent as a function of time for
the D-flip-flop layout.}
\end{figure}

In the run whose population dynamics are shown in figure 3 there are
175 agents in total, distributed initially in a uniform random
distribution over the shared memory. At the start all agents are of
the layer finder type, as the rest of the tasks involved in the layout
extraction process cannot proceed until layer boundaries have been
found. The initial spatial distribution of the agents (receptive
fields) in the environment (or shared memory) was set to a uniform
random distribution.

Very quickly after the start of the program, a large proportion of the
layer finder agents find unlabelled wire boundaries. Thus the
population of layer finders drops precipitously, and the population of
the node labellers (which the layer finders turn into) increases
rapidly. This activity is similar to the action of the immune system
immediately after the infusion of a load of antigen. Antibodies bound
to B-cells which encounter antigen to which they are sensitive
stimulate the B-cells, with the help of T-cells, to produce more
antibodies. Thus there is a rapid increase in the population of
antibodies specific to the antigen shortly after the infusion of
antigen.

The node labeller population begins to decline almost as fast as it
increased, however, as the load balancing effect of the node labeller
dominance strategy comes into action. There are, at the peak of the
node labeller population, many more node labellers than there are
distinct nodes to label. Hence, most of these node labeller agents
will succumb to dominance by other agents.  To speed up labelling of
nodes, the value of the node label written to the shared memory is
either equal to the ID of the agent if no label exists yet, or that of
the current label if it is a higher value than the agent's ID.  When a
node labeller agent, in the course of following the boundary of a
node, encounters a label that is that is the same as the label that it
is currently writing to the environment, it does one of two things,
depending on whether or not the label is equal to its own ID or is
that of another agent. If the label is that of another agent, then the
agent changes into a layer finder. Note that this is safe since the ID
of the current agent has not yet been used to completely label any
wire. If the label is that of the current agent, then the entire
perimeter of the wire has been labelled with that label. The agent
then turns into a node director agent, unless the wire being labelled
is a DIFF wire, in which case the agent turns into a fet labeller
agent.

When the tracing of its DIFF wire is completed a fet labeller agent
changes into a node propagator agent. The reason for having the fet
labeller turn into a node propagator agent rather than a fet output
agent is that there may be more than one transistor on a DIFF wire, so
producing one fet output agent for every DIFF wire will not find all of
the transistors.

Fet labeller agents compete to gain dominance of a transistor
boundary, which they then proceed to label.  Most of the dominated fet
labeller agents will turn into layer finders, to search for wires that
may have been initially shielded by other wires. Thus the layer finder
population rebounds somewhat. This ``second generation'' of layer
finder agents is useful for finding unlabelled nodes which somehow
avoided detection in the first generations of the layer finders.

A small proportion of the
second generation of layer finder agents will turn into node
labellers, but the large proportion will turn into node propagator
agents. Hence, the rise of the population of the node propagator
agents is seen to coincide with the decay of the second generation
layer finders. The fet labeller population rises after an initial
delay, and then decreases as the DIFF regions of the transistors
become marked. The fet output agent population rises after the fet
labeller population does, and decreases very slowly. This slow
decrease is due to the waiting that the fet output agents must do for
the various regions of the transistor to be given stable labels. In
fact, the fet output agents are often the last agents to do useful
work. As the node labeller population decreases the node director
population increases, as the dominant node labellers turn into node
directors when they complete the labelling of their wire. As the node
director population increases, so does the contact finder
population. This is due to the interaction between the messages
written by the node directors and the node propagators, causing the
node propagator agents to turn into contact finders. The contact
finder population stays relatively constant as the contact finders
wait for the overlapping wires under the contacts to have labels
propagated to them via the action of node propagator agents. The
steady state of this algorithm will consist of all node propagator
agents, as all other agents eventually change into node propagators,
and the only transitions away from the node propagator type are to fet
labeller or contact finder types which require conditions that do not
exist once the layout has been completely extracted.

These agent population studies show that complex dynamics arise out of
relatively simple load balancing strategies. Similar dynamics are seen
in the immune system.

\subsection{Algorithm Performance}
How well do the load balancing strategies that we have employed
actually work?  In order to examine the effectiveness of our algorithm
in parallelizing the layout extraction task we need to look at the
relation between the computation time needed to complete the
extraction task and the number of agents used. Such a relation is
shown in figure 4, which shows a log-log plot of completion time
versus the number of agents, for the case of the simple NAND
circuit. The slope of a linear fit to the curve gives the exponent of
the speedup characteristic for the algorithm. In this case the
exponent is about -0.30, which is quite good for a problem like
this. 
\begin{figure}
\begin{center}
\includegraphics[width=5in]{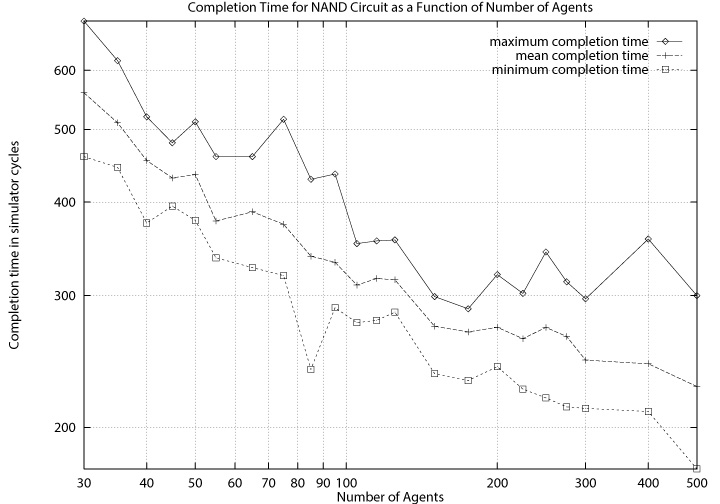}
\end{center}
\caption{The relation between mean task completion times and the
number of agents for the NAND layout extraction problem. The middle
curve is the mean completion time; the top and bottom curves are the
sample minimum and maximum completion times.}
\end{figure}

\section{Summary}
We considered the problem of specifying load balancing strategies for
MIMD systems, which are essential for attaining processing speedup
when multiple processors are used. We examined the approaches that
natural systems, such as the immune system, use to effectively
allocate resources. The approach we took was to abstract the general
strategies used in such natural systems as performing redundancy and
irrelevancy minimization, or as regulatory processes. We presented a
number of heuristic methods for redundancy and irrelevancy detection
and minimization, and related these to processes found in the immune
system. 

Effective local, dynamic, load balancing arises mainly from the
interaction between agents.  The load balancing effect of the
redundancy and irrelevancy minimization heuristics functions primarily
through the modification of an agent's behaviour via messages written
to the environment by other agents. This can be seen in a common form
of interaction in the immune system, where chemical messages
(typically in the form of cytokines) suppress or facilitate the growth
or decline in the population of certain cellular species.  These
heuristics are very simple, but can lead to very efficient load
balancing. Additional research with artificial systems, and further
studies of load balancing strategies employed by natural systems are
needed to refine and expand the above list of approaches.

It should be noted that our approach is by no means restricted to MIMD
computer systems. A collection of mobile robots interacting on a
factory floor \cite{dario,fukuda} is a non-biological exammple of such
a system. Thus, in addition to programming of MIMD parallel computers,
we can use our approach to help design and simulate artificial
systems, such as groups of autonomous robots \cite{beni}, collections
of nanotech assemblers \cite{drexler}, actors and environments in
virtual reality systems, as well as standard computational
applications such as VLSI design and image analysis \cite{clark}.

To illustrate the application of our load balancing techniques, we
presented an algorithm for extracting a circuit netlist from a
graphical representation of a VLSI physical layout. The algorithm was
shown to exhibit respectable speedup characteristics on a simulated
MIMD system, illustrating the effectiveness of the natural-based load
balancing strategies, such as dominance and multi-stage regulation. It
was seen to exhibit complex population dynamics, which arise from the
activity of the load balancing behaviours.

\section{Acknowledgements}
This research was supported in part by the Brown-Harvard-MIT Center
for Intelligent Control Systems, under A.R.O. grant number
DAA103-86-K-0170 and by a grant from the Institute
for Robotics and Intelligent Systems (IRIS), one of the Canadian
Network of Centres of Excellence.

\end{document}